\documentclass[%
 reprint,
nofootinbib,
 amsmath,amssymb,
 aps,
]{revtex4-2}

\usepackage{graphicx}
\usepackage{dcolumn}
\usepackage{bm}
\usepackage{hyperref}

\usepackage{subcaption}

\usepackage{amsmath,amssymb,bm,amsthm,mathrsfs, ascmac, braket,color,stmaryrd}
\usepackage{float}
\usepackage{footmisc}
\usepackage{xcolor}

\usepackage{algorithm2e}
\RestyleAlgo{ruled}

\usepackage{algpseudocode}
\algblock{Input}{EndInput}
\algnotext{EndInput}
\algblock{Output}{EndOutput}
\algnotext{EndOutput}

\begin{document}

\preprint{APS/123-QED}

\title{Quantum decision trees with information entropy}

\author{Zhelun Li}
\email{zhelunli@icepp.s.u-tokyo.ac.jp}
 \author{Koji Terashi}%
 \email{terashi@icepp.s.u-tokyo.ac.jp}

\affiliation{%
International Center for Elementary Particle Physics (ICEPP),\\
The University of Tokyo, 7-3-1 Hongo, Bunkyo-ku, Tokyo 113-0033, Japan
}%

\date{\today}

\begin{abstract}
We present a classification algorithm for quantum states, inspired by decision-tree methods. To adapt the decision-tree framework to the probabilistic nature of quantum measurements, we utilize conditional probabilities to compute information gain, thereby optimizing the measurement scheme. For each measurement shot on an unknown quantum state, the algorithm selects the observable with the highest expected information gain, continuing until convergence. We demonstrate using the simulations that this algorithm effectively identifies quantum states sampled from the Haar random distribution. However, despite not relying on circuit-based quantum neural networks, the algorithm still encounters challenges akin to the barren plateau problem. In the leading order, we show that the information gain is proportional to the variance of the observable's expectation values over candidate states. As the system size increases, this variance, and consequently the information gain, are exponentially suppressed, which poses significant challenges for classifying general Haar-random quantum states. Finally, we apply the quantum decision tree to classify the ground states of various Hamiltonians using physically-motivated observables. On both simulators and quantum computers, the quantum decision tree yields better performances when compared to methods that are not information-optimized. This indicates that the measurement of physically-motivated observables can significantly improve the classification performance, guiding towards the future direction of this approach.
\end{abstract}

\maketitle


\section{Introduction}

The classification of quantum states has been extensively studied in the quantum computing community. With the developments in quantum machine learning (QML)~\cite{Schuld_2014,Biamonte_2017,Mangini_2021,challenges_cerezo}, a wide variety of quantum algorithms have been demonstrated to be effective in this task. However, the progress of quantum neural networks (QNNs)~\cite{Mitarai_2018_paramShift,PhysRevA.101.032308,QCNN_Cong,Wu_2021,Blance_2021,Guan_2021,Cerezo_2021_var} is greatly hindered by the barren plateau phenomenon~\cite{McClean_2018_BP,costFunctionBP_2021,expressiveness_Holmes_2022,Arrasmith_2022,larocca2024review,PRXQuantum.2.040316}, which causes the gradient of a QNN upon random initialization to diminish exponentially as the system size increases. In recent years, several papers investigated various schemes to mitigate the adverse effects of barren plateaus, including shallow circuits~\cite{costFunctionBP_2021,expressiveness_Holmes_2022}, symmetry-preserving ansatz~\cite{Pesah_2021_QCNN_BP,ZLi}, local cost functions~\cite{costFunctionBP_2021}, and many other methods. Recent studies also revealed that most, if not all, existing quantum circuits free of barren plateaus are classically simulable~\cite{cerezo2024doesprovableabsencebarren,bermejo2024quantumconvolutionalneuralnetworks}. This stems from the fact that these algorithms operate within polynomially sized subspaces, enabling classical simulation and negating the need for parameterized quantum circuits. Although parameterized quantum circuits are often considered essential for achieving exponential quantum advantage in machine learning, alternative approaches are being explored. One such approach, proposed in Reference [20], involves using quantum computers exclusively for data acquisition, followed by classical simulation of training losses. This work focuses on developing a QML model based on this hybrid strategy, separating the model into two distinct components: one for acquiring information from measurements on quantum systems, and the other for analyzing this information using classical algorithms.

In this paper, we present an ansatz-free classical algorithm that uses quantum measurement results as inputs. Our algorithm, which we tentatively call a quantum decision tree, is heavily inspired by classical decision tree algorithms such as the Iterative Dichotomiser 3 (ID3)~\cite{ID3_1986}, C4.5~\cite{C45_1992} and classification and regression trees (CART)~\cite{CART1984,randomForest_2001}. The algorithm is designed to correctly label an unknown quantum state from various candidate states by computing the expected information gain in the measurement of possible observables. Since each single quantum measurement only provides a quantized outcome, either $-1$ or $1$ in the computational basis, it is very hard to definitively rule out potential candidate states with just a single measurement.~\footnote{It is possible to rule out a candidate state by just one single measurement if the candidate state is an eigenstate of the measured basis. However, such cases are rare if one is dealing with general quantum states, the expectation values of which are continuously distributed between $-1$ and $1$. } To deal with the probabilistic nature of the quantum theory, we use conditional probabilities to quantitatively update the likelihood of each candidate state after a single measurement. With the definition of likelihood, we can compute the information gain of a particular measurement by calculating the change in the information entropy~\cite{shannon,ID3_1986,C45_1992,CART1984,randomForest_2001}.

In this paper, we discuss the algorithm in its simplest form: an optimized searching algorithm that aims to label an unknown state among various candidates with the fewest number of measurement shots. Since our algorithm uses measurements as inputs, it is fundamentally different from traditional quantum search algorithms, such as Grover's algorithm~\cite{Grover1996AFQ, Nielsen_Chuang_2010,Grover_bound}. We argue that our method is much more robust because it relies solely on the classical treatment of measurement outcomes, rather than on amplitude manipulation techniques~\cite{Grover1996AFQ, Nielsen_Chuang_2010}, which tend to be heavily dependent on hardware and, therefore, are less likely to be realized in the noisy intermediate-scale quantum (NISQ) regime~\cite{Preskill2018quantumcomputingin}. In the numerical experiment sections, we demonstrate the effectiveness of our method by testing it in a searching problem with $20$ Haar-random candidate states in a $10$-qubit system.

Furthermore, we derive the expected information gain per shot and discuss its dependency on the number of qubits. In the leading order, the information gain per shot is proportional to the variance of expectation values in the candidate states. Intuitively, this proves that for Haar random states, we end up with exponentially less information out of each measurement shot with increasing system size, similar to the barren plateau phenomenon in variational quantum algorithms. We claim that barren plateaus are not unique to variational algorithms but are a general phenomenon in all algorithms involving quantum states from large, unstructured quantum systems. To validate our theoretical results, we sampled the information gains in systems of various sizes to observe the exponential suppression. The approximation of information gain using the sample variance is also tested in simulation.

In a broader sense, the working principle of this search algorithm is similar to that of the decision tree and can be easily extended to a classification algorithm that handles unseen data if each class in the training dataset is relatively isolated.  In this paper, we also demonstrate how our quantum decision trees can be used to achieve quantum state classification. We update the probability of the test state being each of the training states in the different classes. Eventually, the algorithm stops when one class has its probability weights summed to be above a certain threshold value. We numerically test the algorithm by attempting to classify the ground states of various Hamiltonians. The performance is further enhanced by choosing physically motivated observables for measurements. The results are obtained from both the simulator and the IBM Kawasaki quantum computer.

The paper is organized as follows. In Section~\ref{sec:info_opt}, we discuss the general structure of our quantum decision tree algorithm.  The general structure of the algorithm, including conditional probability update and information gain computation, is introduced in Section~\ref{ssec:opti_process}. The numerical test of our method is shown in Section~\ref{ssec:classification_numerical}. In Section~\ref{sec:infoGain_large}, we deal with the information gain in large systems. The approximation of information gain in large systems is derived in Section~\ref{ssec:large_approx}, followed by numerical validation in Section~\ref{ssec:large_numerical}.  In Section~\ref{sec:GS_classify}, the quantum decision tree is used to classify ground states of various Hamiltonians on both simulators and the IBM Kawasaki quantum computer. In the end, we discuss the results and their implications in Section~\ref{sec:discussion}.

\section{Information optimization} \label{sec:info_opt}

In this section, we will use classical information theory to investigate the information obtained in each shot of measurement. First, we set up a typical search problem in which we aim to label a test quantum state $| \psi_{t}\rangle$ selected from a set of candidates states $\{| \psi_{i}\rangle\}_{N}$ of size $N$. We then measure the expectation values of an observable set $\{\mathcal{O}_{j}\}$ of size $J$ in all of the candidate states: $\{\langle\mathcal{O}_{j}\rangle_{i}\}$. To label the test state, we pick observables from the set and measure their values in the test state. The resulting measurement outcomes could then be compared with the expectation value of those in the candidate states to update the probability distribution.

\subsection{Optimization process}\label{ssec:opti_process}

\begin{figure*}
  \centering
  \includegraphics[width=\textwidth]{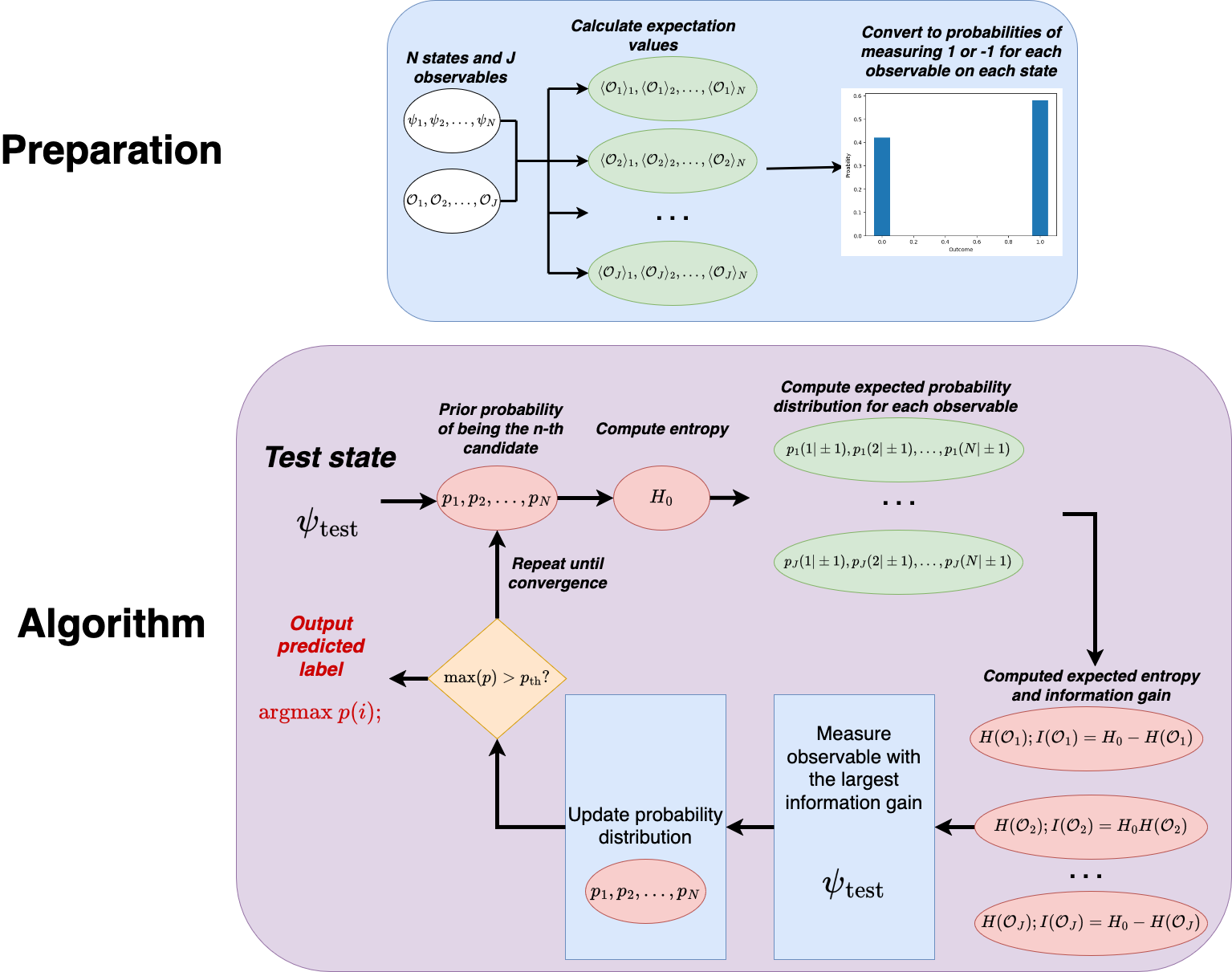}
  \caption{A diagram of the information-optimized quantum algorithm. The probability distribution is updated using conditional probability as shown in Appendix~\ref{sec:Bays}.}
  \label{fig:diagram}
\end{figure*}

Suppose we have decided to use observable $\mathcal{O}_{j}$ to label the test state, we want to know how much information is obtained when a single shot of measurement is taken. We define the probability of the test state being one of the candidates prior to the measurement to be $p(i)$, where $i$ is the index looping through the candidate states. 

After a single shot of measurement, the outcome is either $1$ or $-1$. Since each candidate state has a different expectation value $\langle \mathcal{O}_{j} \rangle_{i}$, their probability of yielding $\pm1$ as the outcome of measuring $\mathcal{O}_{j}$ is different. Using the Bayesian rule, we update the probability of the test state being the i-th candidate state given the outcome:

\begin{equation} \label{eq:Bayes}
p_{j}(i|\pm1) =  \frac{p(i)(1\pm\langle \mathcal{O}_{j} \rangle_{i})}{\sum_{k=0}^{N}p(k)(1\pm\langle \mathcal{O}_{j} \rangle_{k})}.
\end{equation}
The exact derivation of the Bayesian probability update is given in Appendix~\ref{sec:Bays}.

Now that the probability update is determined, we can proceed to calculate the expected information entropy difference. The information entropy is given by $ H = - \sum_{i=0}^{N} p(i) \text{log}_{2}p(i)$. The expected information gain of measuring one shot of observable $\mathcal{O}_{j}$  is defined as the difference between the prior information entropy, $H_{0}$, and the expected information entropy after the measurement, $H(\mathcal{O}_{j})$:
\begin{equation} \label{eq:info}
\begin{split}
I(\mathcal{O}_{j}) &= H_{0} - H(\mathcal{O}_{j}),\\
\end{split}
\end{equation}
where $H(\mathcal{O}_{j}) = (p(1)H(1) + p(-1)H(-1))$ with $H(\pm1) =  - \sum_{i=0}^{N} p_{j}(i|\pm1) \text{log}_{2}p_{j}(i|\pm1)$.

The information-optimized strategy is then given by the following:
\begin{enumerate}
    \item Start with a probability distribution $p(i)$
    \item Compute the expected information gain, $I(\mathcal{O}_{j})$, for all available observables, $\{ \langle \mathcal{O}_{j} \rangle\}$.
    \item Pick the observable with the largest information gain and take one shot of measurement.
    \item Update the probability distribution using the Bayesian rule and repeat the procedures from step 1 unless one candidate's probability reaches the desired confidence level (CL) threshold $p_{\text{th}}$.
\end{enumerate}
A schematic diagram of the strategy is presented in Figure~\ref{fig:diagram}, while its algorithmic representation is given in Algorithm~\ref{QDT}.

\begin{algorithm}[H]
  \caption{Quantum Decision Tree}
  \label{QDT}

  \SetKwInOut{Input}{Input}
  \SetKwInOut{Output}{Output}
  
  \Input{$p(i) = \frac{1}{N}$ \tcp*{Initial probability}}

  \While{$\max(p(i)) < p_{\text{th}}$}{
      \For{$j \gets 1$ \textbf{to} $J$}{
          Compute $I(\mathcal{O}_{j})$
      }
      Measure $\operatorname{argmax} I(\mathcal{O}_{j})$
      
      Update $p(i)$ using conditional probability
  }
  
  \Output{$\operatorname{argmax} p(i)$ \tcp*{Predicted label}}
  
\end{algorithm}

\subsection{Quantum state searching}\label{ssec:classification_numerical}
With the method defined in the last subsection, we proceed to demonstrate its effectiveness in a typical searching task.

\begin{figure*}[t]
  \centering
  \includegraphics[width=0.8\textwidth]{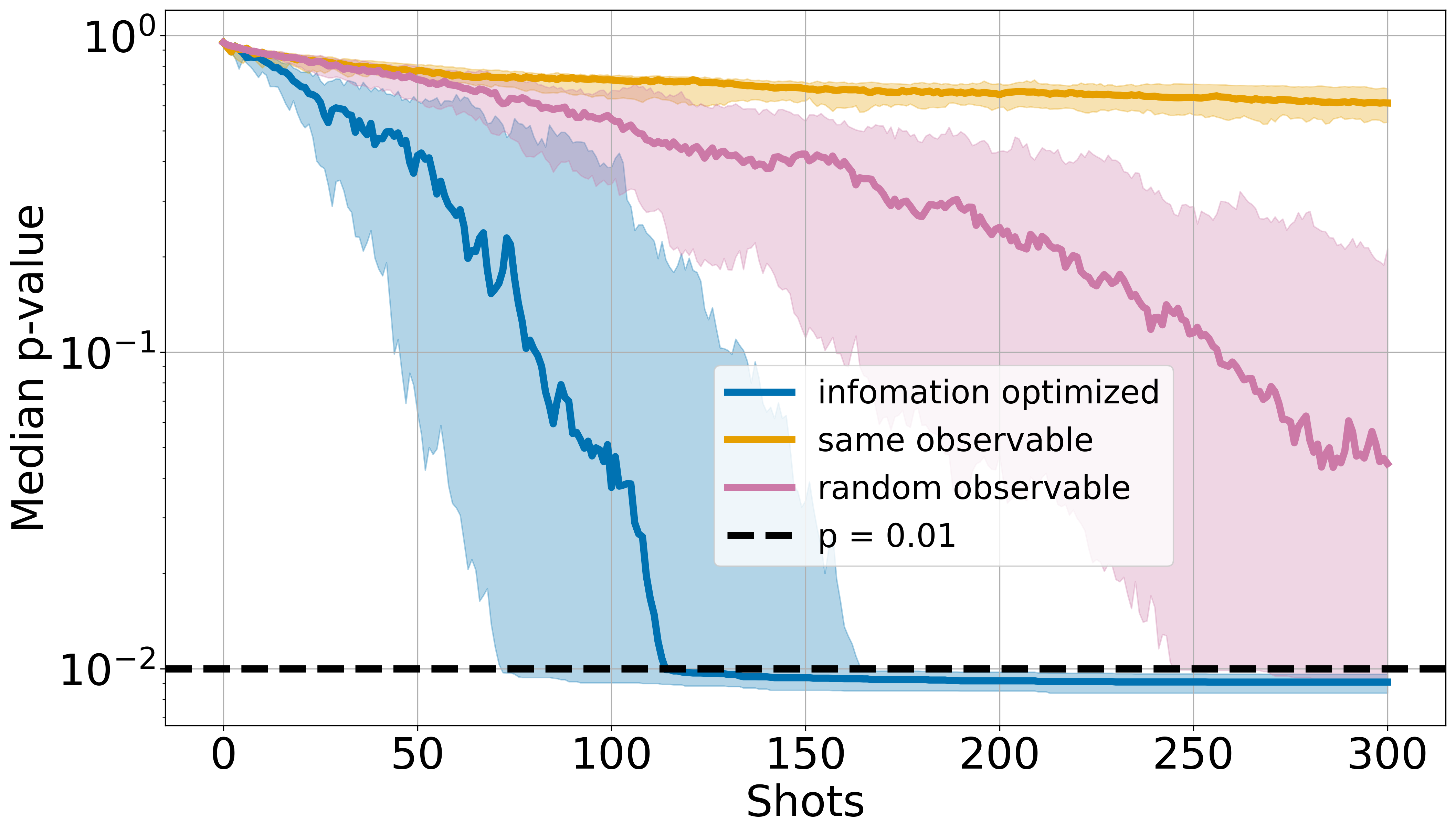}
  \caption{The median p-value of the quantum state classification against the number of shots taken. The error band is computed at $25\%$ and $75\%$ quantiles. The algorithm terminates after the p-value drops below $0.01$.}
  \label{fig:median_p_value_sim_log}
\end{figure*}

In a simulated $10$-qubit system, we prepared $20$ random quantum states and $20$ observables for the task.  To generate random quantum states, we use a sufficiently deep quantum circuit with random initialization~\cite{Sim2019ExpressibilityAE,expressibility_nakaji}. Our task is to label a randomly drawn test state at $99\%$ CL within $300$ shots of measurements. Equivalently, the algorithm stops if the probability of a given candidate state exceeds $99\%$, which indicates a p-value of $0.01$. We compare the performances of three methods: 
\begin{enumerate}
    \item Measuring the same observable for as many times as needed. The observable is chosen to be the one that gives the highest information gain in the first iteration.
    
    \item Randomly pick an observable for each shot of measurement.
    
    \item Information-optimized measurement: picking the observable with the highest expected information gain given the current probability distribution.
\end{enumerate}
The experiment runs for $100$ test cases, the result of which is shown in Figure~\ref{fig:median_p_value_sim_log}. The median p-value is plotted against the number of shots taken, while the error bands correspond to $25\%$ and $75\%$ quantiles, respectively. The  results show that the information-optimized algorithm converges to $99\%$ CL much faster than the other two alternatives. On average, our quantum decision tree with the information optimization typically converges within $150$ shots of measurements.

\section{Expected information gain in large systems}\label{sec:infoGain_large}
In this section, we will investigate the scaling of the expected information gain per shot with respect to the number of qubits. The task is to classify the test state $| \psi \rangle$ among $N$ Haar random state candidates $\{ \psi_{i}\}$~\cite{Haar_Mele_2024}, silimar to what was shown in Section~\ref{sec:info_opt}.  We will derive the approximated information gain that depends on the system size and observe that the expected information gain is proportional to the sample variance of the expectation value of the chosen observable. For Haar random states, this eventually results in the exponential suppression of the expected information gain per shot.

\subsection{Approximations of expected information gain}\label{ssec:large_approx}

In this section, we will investigate the information gain in the measurement of an arbitrary observable $\mathcal{O}$. Before any measurement, the prior probability distribution is uniform across the whole set of candidate states, i.e., $p_{0} = \frac{1}{N}$. The entropy is therefore:
\begin{equation} \label{eq:H0}
\begin{split}
H_{0} &= -\sum_{i=0}^{N} p_{0}\text{log}_{2}p_{0}\\
&= -\sum_{i=0}^{N} \frac{1}{N}\text{log}_{2}\frac{1}{N}\\
&= \text{log}_{2}N.
\end{split}
\end{equation}

Given an observable $\mathcal{O}$, we first measure its expectation values across the whole set of candidate states: $\langle \mathcal{O} \rangle_{i}$. Now we proceed to measure the expected entropy after measuring one shot of $\mathcal{O}$:

\begin{equation} \label{eq:H}
\begin{split}
H' &= p(1)H(1) + p(-1)H(-1)\\
       &\approx H_{0} -\frac{V[\langle \mathcal{O} \rangle]}{2\text{ln}2}-\frac{E[\langle \mathcal{O} \rangle]^2}{2\text{ln}2}(1-\frac{2}{N}),
\end{split}
\end{equation}
where $V[\langle \mathcal{O} \rangle]$ and $E[\langle \mathcal{O} \rangle]$ are the variance and the mean value of $\langle \mathcal{O} \rangle$ across the whole set of candidate states:

\begin{equation} \label{eq:ev}
\begin{split}
E[\langle \mathcal{O} \rangle]&=\sum_{i=0}^{N}\frac{\langle\mathcal{O}\rangle_{i}}{N}\\
V[\langle \mathcal{O} \rangle]&=\sum_{i=0}^{N}\frac{\langle\mathcal{O}\rangle_{i}^{2}}{N} - (E[\langle \mathcal{O} \rangle])^{2}.
\end{split}
\end{equation}

In the derivation of Equation~\ref{eq:H}, we approximate the entropy by taking Taylor expansion to the second order. The full derivation is given in Appendix~\ref{sec:entropyDerivation}.  The expected information gain per shot is then given by:

\begin{equation} \label{eq:I}
\begin{split}
I &= H_{0} -H'\\
&\approx \frac{V[\langle \mathcal{O} \rangle]}{2\text{ln}2}+\frac{E[\langle \mathcal{O} \rangle]^2}{2\text{ln}2}(1-\frac{2}{N}).
\end{split}
\end{equation}

The expected information gain shown in Equation~\ref{eq:I} is strictly positive for $N>1$. This is expected since we started with the uniform probability distribution that yields maximum entropy. In other words, all measurements will yield non-negative information gain since the resulting entropy cannot be higher than the initial one.

Following the recipe in Reference~\cite{McClean_2018_BP,ipsen2015productsindependentgaussianrandom}, we proceed to calculate $V[\langle \mathcal{O} \rangle]$ and $E[\langle \mathcal{O} \rangle]$ over Haar random states:

\begin{equation} \label{eq:E}
\begin{split}
E[\langle \mathcal{O} \rangle] &= \int d\mu(U) \langle 0 | U^{\dagger} \mathcal{O} U | 0 \rangle\\
&=\frac{\text{Tr}(\mathcal{O})}{2^{n}},
\end{split}
\end{equation}

\begin{equation} \label{eq:V}
\begin{split}
V[\langle \mathcal{O} \rangle] &= E [ \langle \mathcal{O}^{2} \rangle] - E[\langle  \mathcal{O} \rangle]^{2} \\
&\approx \frac{\text{Tr}(\mathcal{O}^{2})}{2^{2n}-1}  - \frac{\text{Tr}^{2}(\mathcal{O})}{2^{2n}},
\end{split}
\end{equation}
where $ n $ is the number of qubits in the system.

Both $V[\langle \mathcal{O} \rangle]$ and $E[\langle \mathcal{O} \rangle]$ are exponentially suppressed over the Haar random states. Plugging these into Equation~\ref{eq:I}, we naturally conclude that the expected information gain per shot is decreasing exponentially with respect to the number of qubits. 

In the case that we use Pauli operators, which are by definition traceless,  $E[\langle \mathcal{O} \rangle]$ becomes zero. And then the information gain per shot is therefore proportional to the variance:

\begin{equation} \label{eq:I_Vonly}
\begin{split}
I \approx \frac{V[\langle \mathcal{O} \rangle]}{2\text{ln}2}.
\end{split}
\end{equation}

When computing the expectation value of the information gain in the sample, $\mathbb{E}\left[I_{s}\right]$, we note the definition of the variance in Equation~\ref{eq:ev} is biased.  This biased definition of variance is used in the whole derivation, introducing a non-negligible sampling bias when $N$ is small. Instead, the unbiased sample variance, $V_s[\langle \mathcal{O} \rangle]$, should be used, which is defined as:

\begin{equation} \label{eq:sampleVariance}
\begin{split}
V_s[\langle \mathcal{O} \rangle] = \frac{\sum_{i=0}^{N} (\langle\mathcal{O}\rangle_{i} - E_s[\langle \mathcal{O} \rangle])^{2}}{N-1},
\end{split}
\end{equation}
where $E_s[\langle \mathcal{O} \rangle]$ is the sample mean of the observable's expectation value.

The unbiased sample variance definition satisfies  $\mathbb{E} \left[ V_s[\langle \mathcal{O} \rangle]   \right] = V^{\text{Haar}}[\langle \mathcal{O} \rangle]$, where $V^{\text{Haar}}[\langle \mathcal{O} \rangle]$ is the variance of $\langle \mathcal{O} \rangle$ over the entire Haar-random states. Therefore, we see that an extra factor is introduced in the expectation value $\mathbb{E}\left[V[\langle \mathcal{O} \rangle]\right] $:

\begin{equation} \label{eq:bias}
\begin{split}
\mathbb{E}\left[V[\langle \mathcal{O} \rangle]\right] &= \mathbb{E} \left[ \frac{\sum_{i=0}^{N} (\langle\mathcal{O}\rangle_{i} - E_s[\langle \mathcal{O} \rangle])^{2}}{N}  \right]\\
&= \mathbb{E} \left[ \frac{N-1}{N} V_s[\langle \mathcal{O} \rangle]   \right]\\
&=\frac{N-1}{N}  V^{\text{Haar}}[\langle \mathcal{O} \rangle].
\end{split}
\end{equation}

As a result, we derive the sample information gain $I_{s}$ with a dependence on $N$:
\begin{equation} \label{eq:I_Vonly_sample}
\begin{split}
\mathbb{E}\left[I_{s}\right] \approx \frac{V^{\text{Haar}}[\langle \mathcal{O} \rangle]}{2\text{ln}2} (1-\frac{1}{N}).
\end{split}
\end{equation}

From the above derivation, we observe that the information obtained from measuring an observable is roughly proportional to the variance of its expectation value across the set of candidate states. Intuitively, higher variance means the observable yields more distinct values for different candidate states, leading to more information obtained per shot.

\begin{figure*}
\captionsetup[subfigure]{justification=centering}
     \centering
     \begin{subfigure}[b]{0.8\textwidth}
         \centering
         \includegraphics[width=\textwidth]{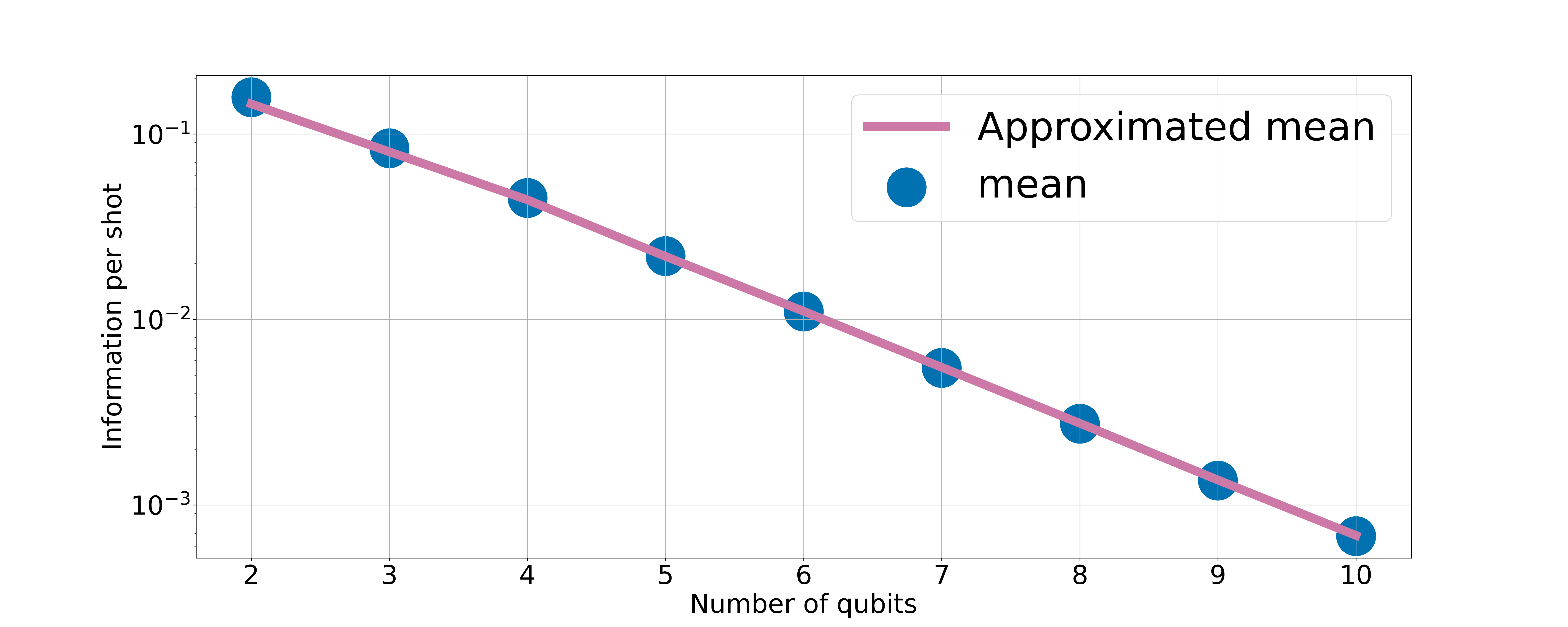}
         \caption{Information against the number of qubits.}
         \label{fig:Info_mean}
     \end{subfigure}
     \bigskip
     \begin{subfigure}[b]{0.8\textwidth}
         \centering
         \includegraphics[width=\textwidth]{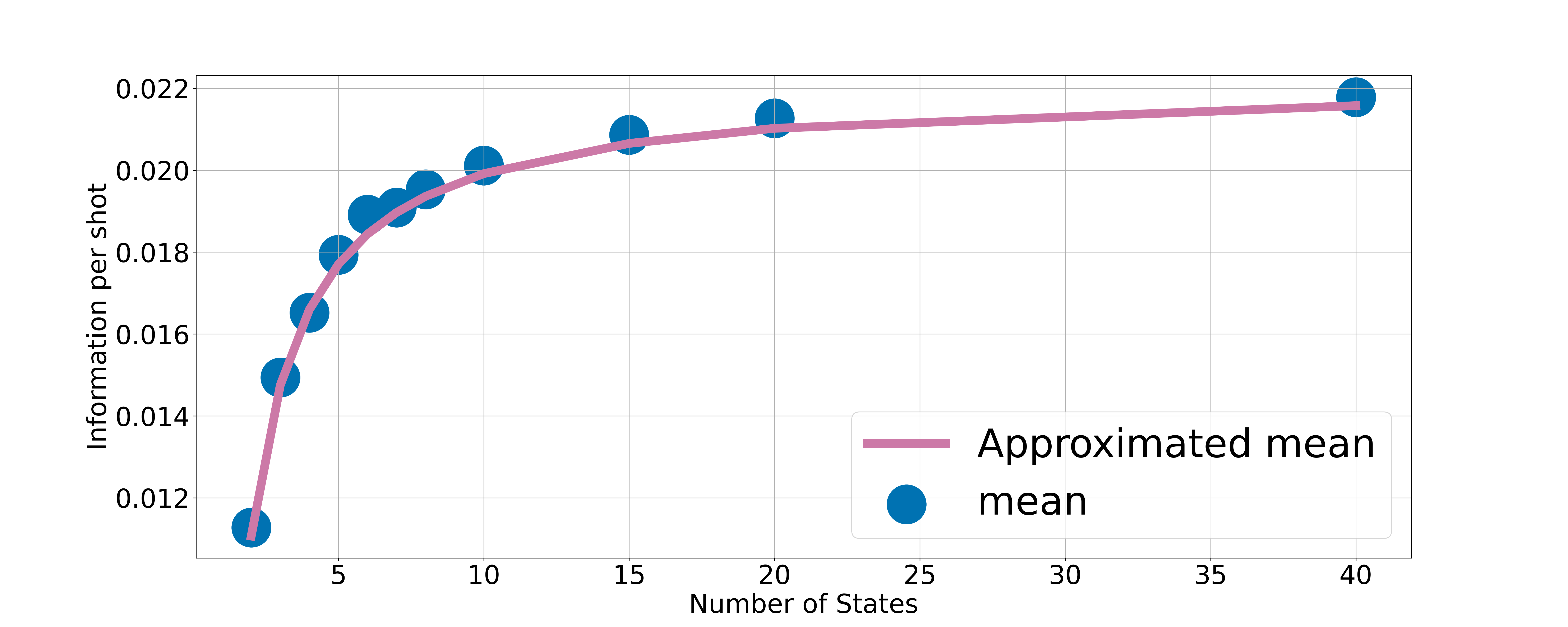}
         \caption{Information against the number of candidate states in a $5$-qubit system.}
         \label{fig:info_sample}
     \end{subfigure}

        \caption{The mean information per shot against the number of qubits (a) and the number of states (b). In both cases, the approximations are calculated using the variances of the expectation values. Figure~\ref{fig:Info_mean} shows the average information per shot over a set of $100$ states and $100$ observables. Figure~\ref{fig:info_sample} shows the mean information per shot in a $5$-qubits system with $200$ observables. The approximated information in Figure~\ref{fig:info_sample} is obtained by first computing $V^{\text{Haar}}[\langle \mathcal{O} \rangle]$ with large numbers of states and then introducing the factor of $(1-\frac{1}{N})$ in Equation~\ref{eq:I_Vonly_sample}. }
        \label{fig:info_mean_sample}
\end{figure*}

\subsection{Numerical studies}\label{ssec:large_numerical}
To experimentally verify the theoretical conclusions from the last subsection, we numerically simulate the quantum states using circuits constructed by Qiskit~\cite{Qiskit}. We used deep circuits and random initializations to produce quasi-Haar-random states for our analysis.

First, we attempt to verify Equation~\ref{eq:I_Vonly_sample} by computing information per shot with varying the number of qubits. For each qubit number, we generate $100$ Haar-random states to be distinguished via measuring a set of $100$ randomly generated observables. As before, we start with the uniform prior probability over the $100$ states and calculate information per shot exactly using Equation~\ref{eq:info}. In the end, we average the information per shot over all observables to get the mean information per shot for a particular qubit number. Alternatively, we also compute the approximated information per shot according to Equation~\ref{eq:I_Vonly_sample} by evaluating the variance of the expectation values of the observables across the set of Haar-random states, $V^{\text{Haar}}[\langle \mathcal{O} \rangle]$, which is subsequently plugged into Equation~\ref{eq:I_Vonly_sample}. The calculations were made for numbers of qubits between $2$ and $10$ with the results shown in Figure~\ref{fig:Info_mean}. We see that Equation~\ref{eq:I_Vonly_sample} reliably approximates the exponentially diminishing information per shot over the entire qubit number range.

We then proceed to verify the bias factor in Equation~\ref{eq:I_Vonly_sample} by fixing the number of qubits at 5 and computing the mean information over $200$ observables on candidate state sets of various sizes. The results of the mean information per shot in sets with the number of states between $2$ and $40$ are shown in Figure~\ref{fig:info_sample}. The approximated mean uses Equation~\ref{eq:I_Vonly_sample} with the variance over the entire set of Haar-random states. We see that Equation~\ref{eq:I_Vonly_sample} accurately captures the sampling bias in the mean information per shot in small sets.

\section{Ground state classifications}\label{sec:GS_classify}

\begin{figure*}[t]
  \centering
  \includegraphics[width=0.8\textwidth]{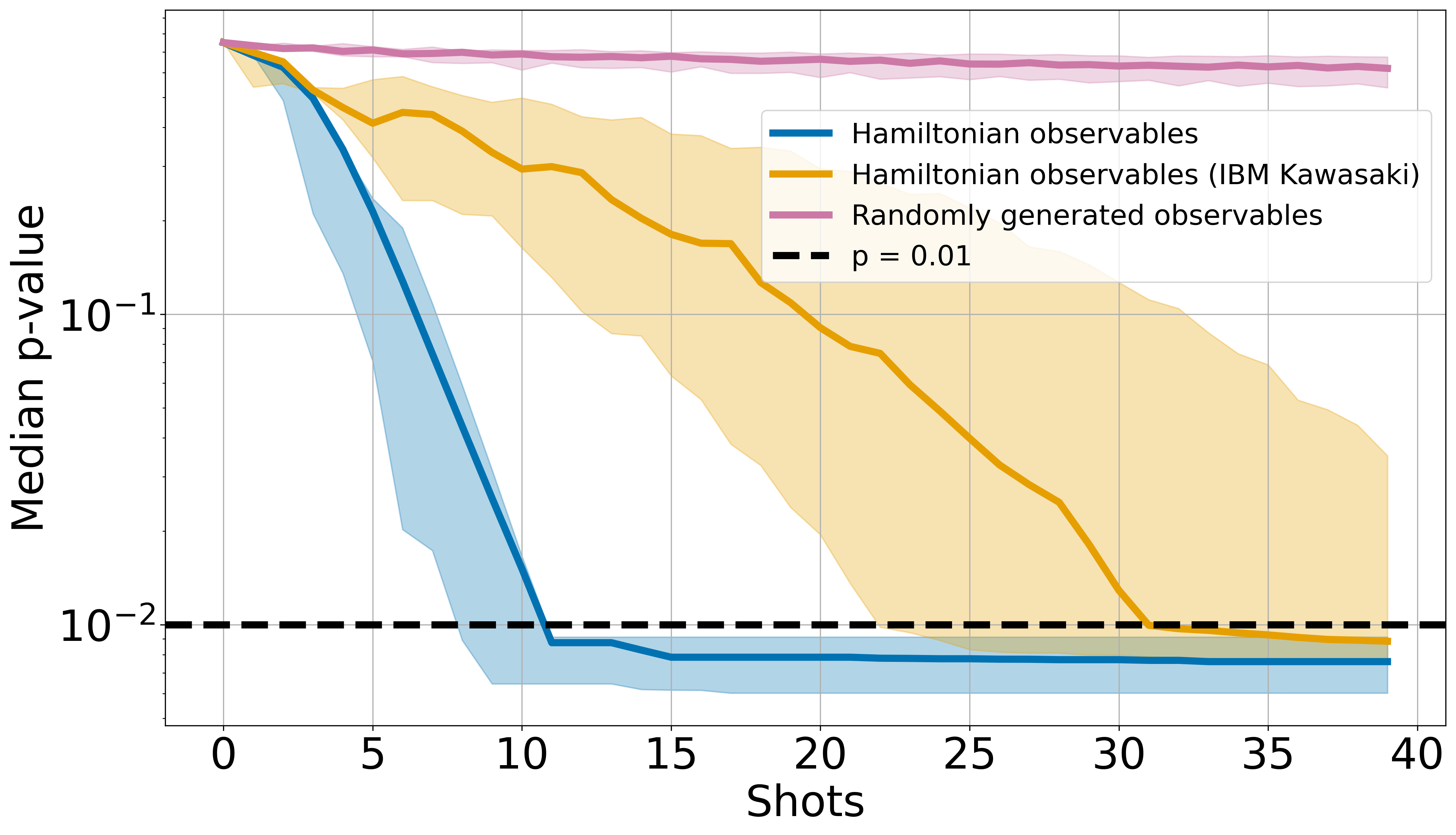}
  \caption{The median p-value of the Hamiltonian ground state classification against the number of shots taken. The error band is computed at $25\%$ and $75\%$ quantiles. The blue curve is obtained by using the observables that are found in the set of Hamiltonians.  The yellow curve is obtained by computing the Hamiltonian observables on the IBM Kawasaki quantum computer. The purple curve is obtained by using the best observable in randomly generated sets of Pauli operators on the simulator.  The algorithm terminates after the p-value drops below $0.01$.}
  \label{fig:median_p_value_sim_log_GS}
\end{figure*}

In this section, we will adapt our quantum decision tree to handle classification tasks. First, we see that the method in Section~\ref{ssec:opti_process} updates the probabilities of the test state being each of the candidates. Assuming that the candidate states are representative of their respective classes, the probability of the test state belonging to a particular class can be determined by summing the probabilities of the test state corresponding to any of the training states within that class. Given $n_c$ distinct classes, each containing $n_{s}$ training states, the probability of the test state being classified as class c is computed as:

\begin{equation} \label{eq:class}
\begin{split}
p_{\text{class}}(c) = \sum_{i=0}^{n_s} p(i,c).
\end{split}
\end{equation}

Since our ultimate task is to determine the label of class, we compute the entropy using the class probability distribution and seek to maximize the information gain in the entropy of classes. The entropy with respect to the classes is given by:

\begin{equation} \label{eq:H_class}
\begin{split}
H_{\text{class}} &= -\sum_{c = 0}^{n_{c}} p_{\text{class}}(c) \text{log}_{2} p_{\text{class}}(c).
\end{split}
\end{equation}

    It is easy to see that once the entropy gets properly defined, we can proceed to use the methodology in Section~\ref{ssec:opti_process} to find the best observable that yields the maximum information gain. As a numerical test, we compute the ground states of four classes of Hamiltonians, each with $100$ sets of parameters. We picked $75$ states in each class to be the ``training set'', where the likelihood is evaluated, and the remaining $25$ states in each class to be the test states, which are labeled using our algorithm. For this task, we use four physically-motivated Hamiltonians: the Heisenberg model~\cite{Heis,Heis2}, spin-$1/2$ chain with symmetry-protected topological (SPT) phase~\cite{QCNN_Cong,deGroot2022symmetryprotected,bermejo2024quantumconvolutionalneuralnetworks,SPTQC}, the transverse-field Ising model~\cite{utility}, and a synthetically produced Hamiltonian which we call the XYZ Hamiltonian. The Hamiltonians and their parameter ranges are listed below:

\begin{enumerate}
    \item Heisenberg:
        \begin{itemize}
            \item $H_{\text{Heis}} = -\sum_{j=0}^{N}\sigma_{j}\sigma_{j+1}-h\sum_{j=0}^{N}\sigma_{j}$, where $\sigma$ denotes the Pauli X, Y, Z.
            \item Range of parameters: $h$ between $1.1$ and $3.1$ in intervals of $0.02$.
        \end{itemize}

    \item SPT:
        \begin{itemize}
            \item $H_{\text{SPT}} = -\sum_{j=0}^{N-2}Z_{j}X_{j+1}Z_{j+2} - h_{1}\sum_{j=0}^{N}X_{j} -h_{2}\sum_{j=0}^{N-1}X_{j}X_{j+1}$ 
            \item Range of parameters: $h_{1}$ between $0$ and $0.3$ in intervals of $0.06$, $h_{2}$ between $-0.2$ and $0.2$ in intervals of $0.02$
        \end{itemize}

    \item Transverse-field Ising:
        \begin{itemize}
            \item $H_{\text{Ising}} = -\sum_{j=0}^{N-1}Z_{j}Z_{j+1} - h\sum_{j=0}^{N}X_{j}$ 
            \item Range of parameters: $h$ between $1.1$ and $3.1$ in intervals of $0.02$.
        \end{itemize}

    \item XYZ:
        \begin{itemize}
            \item $H_{\text{XYZ}} = -\sum_{j=0}^{N-2}X_{j}Y_{j+1}Z_{j+1} - h\sum_{j=0}^{N}A_{j}$, where $A$ alternates between the three Pauli operators in the order of $X, Y, Z$. 
            \item Range of parameters: $h$ between $1.1$ and $3.1$ in intervals of $0.02$.
        \end{itemize}
\end{enumerate}

For each of the Hamiltonians listed above, we compute the ground state using variational quantum eigensolver on a simulator with $10$ qubits.  In this task, since we are dealing with the ground states, it is physically motivated to use observables found in the Hamiltonians. In Section~\ref{sec:infoGain_large}, we showed that as the system size grows, most observables deliver exponentially diminishing information. By contrast, using Hamiltonian observables exploits the fact that the ground state is more likely to take extreme values in them. Overall, there are $73$ distinct observables that appeared in all four Hamiltonians on a $10$ qubits system. In comparison, we also execute our algorithm with a set of $73$ randomly generated observables by drawing uniformly from all possible combinations of Pauli operators. After the ground states are obtained, we then compute the expectation values of specifically chosen observables used in our algorithms. In this numerical test, we compute the expectation values on both the simulator and the IBM Kawasaki quantum computer. Since it is much less efficient to obtain single-shot measurements on IBM Kawasaki, we measure the estimated mean value and its error on the IBM machine and then use these to simulate the single-shot results offline. To mitigate the noise on IBM Kawasaki, we implement randomized Pauli twirling~\cite{Pauli} and zero noise extrapolation~\cite{ZNE,Error_mitigation} using the IBM qiskit package.

The results of the numerical experiment are shown in Figure~\ref{fig:median_p_value_sim_log_GS}, where the quantum decision tree classifier yields a faster convergence when it uses the Hamiltonian observables. Since the IBM Kawasaki quantum computer introduces hardware noises, its performance significantly deteriorates. These results confirm our intuition: the observables in the Hamiltonian are information-dense since the ground states are by definition states that yield extreme values in the full Hamiltonian. To the eyes of the other observables not appearing in the Hamiltonian, the ground states are no different from any other Haar-random states, in which the information per shot is greatly suppressed. Therefore, this leads to a significantly poor performance in classifying the ground states when the random observables are used.

\section{Discussion and conclusions}\label{sec:discussion}

In this paper, we introduced an algorithm to label unknown quantum states using information gain. Our algorithm chooses the optimal observable, which yields the highest expected information gain, to measure and subsequently updates the likelihoods of each candidate state using conditional probabilities based on their expectation values of the chosen observable and the outcome of the measurement. Our information-optimized algorithm is effective and efficient as shown in Figure~\ref{fig:median_p_value_sim_log} and Figure~\ref{fig:median_p_value_sim_log_GS}.  However, further investigations reveal that the expected information gain of a particular observable is proportional to the variance of the candidate states' expectation values.  This introduces an exponential suppression of information gain as the system size increases since the expectation values of observables over general Haar-random quantum states are diminishing exponentially. These results are in good agreement with numerical simulation shown in Figure~\ref{fig:info_mean_sample}. By computing the inter-class entropy, we are able to adapt the algorithm to perform classifications. As shown in Figure~\ref{fig:median_p_value_sim_log_GS}, we observe that our algorithm yields good performance on both the simulator and the IBM Kawasaki quantum computer when applied to ground state classification tasks.

One of the main advantages of our method is its robustness. Unlike QNNs~\cite{Mitarai_2018_paramShift,PhysRevA.101.032308,QCNN_Cong} that require a quantum circuit to prepare the states and a parameterized quantum circuit to optimize them, our method needs only the former for the classification. As a result, it is less affected by the noisy quantum gates and entanglement operations in NISQ machines. However, the noise in quantum hardware is certainly a non-negligible factor when applying this method.  When determining the expectation values of the observables in each candidate state, the real quantum machines could only estimate the values up to the hardware resolutions. As shown in Figure~\ref{fig:median_p_value_sim_log_GS}, the hardware noise has a non-negligible negative impact on the algorithm performance.  In the limit of large system size, the increasing hardware noise in NISQ machines will smear out the information content in the expectation values. An interesting future direction of study will be the development of a good noise model on the information gain such that the responses from real quantum machines can be studied quantitatively.

Aside from the study shown in this paper, it is also crucial to understand how our method could be extended to physically-motivated problems.  If the quantum states to be classified are known to have certain properties, one might be able to use this information to cherry-pick the information-dense observables. A clear example is the classification problem in Section~\ref{sec:GS_classify}, where the classes of states are relatively concentrated in subspaces represented by the problem Hamiltonian. As a result, the Hamiltonian observables are shown to give much higher information gain compared to randomly chosen observable sets. This exploits the fact that ground states are by definition states that take extreme values in the operators of their Hamiltonians and thus deviate from the exponentially diminishing expectation values of Haar-random states. We believe that such exploitation could offer a significant increase in the information gain and eventually allow us to deal with large-scale quantum state classifications in a problem-specific fashion.

\vspace*{0.5cm}
\acknowledgments
We would like to thank Dr.~Lento Nagano for the discussions and his valuable opinions.  This work is supported by the Center of Innovations for Sustainable Quantum AI (Japan Science and Technology Agency Grant Number JPMJPF2221). We acknowledge the usage of IBM quantum computers for this work.

\onecolumngrid
\appendix

\section{Bayesian probability update}\label{sec:Bays}
The expectation value of $\mathcal{O}$ for a state $|\psi_{i}\rangle$ could be expressed using its probability of obtaining $1$, $p(1|i)$, as the outcome of measuring $\mathcal{O}$:

\begin{equation} \label{eq:EO}
\begin{split}
\langle \mathcal{O} \rangle_{i} &= \frac{p(1|i) - (1-p(1|i))}{2} = \frac{1-2p(1|i)}{2}.
\end{split}
\end{equation}

Therefore, the probability of measuring $\pm1$ in this state, $p_{i}(\pm1)$, is:

\begin{equation} \label{eq:p1}
\begin{split}
p(\pm1|i) &= \frac{1\pm\langle \mathcal{O} \rangle_{i}}{2}.
\end{split}
\end{equation}

Using the Bayesian rule, the probability of the test state being the i-th candidate state after measuring $\pm1$ as the outcome of $\mathcal{O}$ is given by:

\begin{equation} \label{eq:bays_derivation}
\begin{split}
p(i|\pm1)= p(i) \frac{p(\pm1|i)}{p(\pm1)} = p(i) \frac{(1\pm\langle \mathcal{O} \rangle_{i})/2}{\sum_{k=0}^{N}p(k)(1\pm\langle \mathcal{O} \rangle_{k})/2} =  \frac{p(i)(1\pm\langle \mathcal{O} \rangle_{i})}{\sum_{k=0}^{N}p(k)(1\pm\langle \mathcal{O} \rangle_{k})}.
\end{split}
\end{equation}

\section{Approximating expected information gain per shot}\label{sec:entropyDerivation}
In this Appendix, we will denote $V[\langle \mathcal{O} \rangle]$ and $E[\langle \mathcal{O} \rangle]$ as $V$ and $E$ in short.  In the set of $N$ candidate states, these values are defined as:

\begin{equation} 
\begin{split}
E&=\sum_{i=0}^{N}\frac{\langle\mathcal{O}\rangle_{i}}{N},\\
V&=\sum_{i=0}^{N}\frac{\langle\mathcal{O}\rangle_{i}^{2}}{N} - E^{2}.
\end{split}
\end{equation}
Since we have taken these candidates to be distributed according to the Haar random distribution, these values will approach the ones derived in Equation~\ref{eq:E} and Equation~\ref{eq:V} as $N$ increases.

First, we examine the approximation of the logarithm term in the expected information gain. When $\langle\mathcal{O}\rangle$ is small, we perform the Taylor expansion:

\begin{equation} \label{eq:log_approx}
\begin{split}
\text{log}_{2}(1+\langle\mathcal{O}\rangle) = \frac{\langle\mathcal{O}\rangle}{\text{ln}2}-\frac{\langle\mathcal{O}\rangle^{2}}{2\text{ln}2} +O(\langle\mathcal{O}\rangle^{3}).
\end{split}
\end{equation}

The prior probability of measuring $\pm1$ as the outcome of $\mathcal{O}$ is given by:

\begin{equation} \label{eq:p10}
\begin{split}
p(\pm1) &= \sum_{i=0}^{N}p_{0} \cdot p^{i}(1) \\
&=  \sum_{i=0}^{N}\frac{(1+\langle\mathcal{O}\rangle_{i})/2}{N}\\
&=\frac{N(1+E)}{2}.
\end{split}
\end{equation}

The entropy after measuring $1$ as the outcome is:

\begin{equation} \label{eq:H1}
\begin{split}
H(1) &= - \sum_{i=0}^{N} \frac{p^{i}(1)}{p(1)} \text{log}_{2} \frac{p^{i}(1)}{p(1)}\\
&= -\frac{1}{2p(1)} \sum_{i=0}^{N}(1+\langle\mathcal{O}\rangle_{i})\text{log}_{2} \frac{(1+\langle\mathcal{O}\rangle_{i})}{N(1+E)}\\
&= -\frac{1}{2p(1)} \sum_{i=0}^{N}(1+\langle\mathcal{O}\rangle_{i})[\text{log}_{2}(1+\langle\mathcal{O}\rangle_{i}) -\text{log}_{2}N -\text{log}_{2} (1+E)]\\
&\approx -\frac{1}{2p(1)} \sum_{i=0}^{N}(1+\langle\mathcal{O}\rangle_{i})\left[\frac{\langle\mathcal{O}\rangle_{i}}{\text{ln}2} - \frac{\langle\mathcal{O}\rangle_{i}^2}{2\text{ln}2} -H_{0}-\frac{E}{\text{ln}2} +\frac{E^{2}}{2\text{ln}2}\right]\\
&= -\frac{1}{2p(1)} \left(\sum_{i=0}^{N}\frac{\langle\mathcal{O}\rangle_{i}}{\text{ln}2} - \sum_{i=0}^{N}\frac{\langle\mathcal{O}\rangle_{i}^{2}}{2\text{ln}2} 
+\sum_{i=0}^{N}\frac{\langle\mathcal{O}\rangle_{i}^{2}}{\text{ln}2}
-\sum_{i=0}^{N}\frac{\langle\mathcal{O}\rangle_{i}^{3}}{2\text{ln}2}
 + \sum_{i=0}^{N}(1+\langle\mathcal{O}\rangle_{i}) \left[ -H_{0}-\frac{E}{\text{ln}2} +\frac{E^{2}}{2\text{ln}2}\right]\right)\\
 &\approx  -\frac{1}{2p(1)} \left(\sum_{i=0}^{N}\frac{\langle\mathcal{O}\rangle_{i}}{\text{ln}2} + \sum_{i=0}^{N}\frac{\langle\mathcal{O}\rangle_{i}^{2}}{2\text{ln}2} -\sum_{i=0}^{N}\frac{\langle\mathcal{O}\rangle_{i}^{3}}{2\text{ln}2}
+ \sum_{i=0}^{N}(1+\langle\mathcal{O}\rangle_{i}) \left[ -H_{0}-\frac{E}{\text{ln}2} +\frac{E^{2}}{2\text{ln}2}\right]\right)\\
&= -\frac{1}{2p(1)} \left(   \frac{E}{\text{ln}2} +\frac{N(V+E^{2})}{2\text{ln}2} -\sum_{i=0}^{N}\frac{\langle\mathcal{O}\rangle_{i}^{3}}{2\text{ln}2} -(N+E)H_{0} - \frac{E}{\text{ln}2} - \frac{E^{2}}{2\text{ln}2} +\frac{E^{3}}{2\text{ln}2} \right).
\end{split}
\end{equation}

Similarly, we can work out the entropy after measuring $-1$ as the outcome:

\begin{equation} \label{eq:H-1}
\begin{split}
H(-1) &= - \sum_{i=0}^{N} \frac{p^{i}(-1)}{p(-1)} \text{log}_{2} \frac{p^{i}(-1)}{p(-1)}\\
&\approx -\frac{1}{2p(-1)} \left( -\frac{E}{\text{ln}2} +\frac{N(V+E^{2})}{2\text{ln}2} +\sum_{i=0}^{N}\frac{\langle\mathcal{O}\rangle_{i}^{3}}{2\text{ln}2} -(N-E)H_{0} + \frac{E}{\text{ln}2} - \frac{E^{2}}{2\text{ln}2} -\frac{E^{3}}{2\text{ln}2} \right).
\end{split}
\end{equation}

The overall expected entropy is calculated by summing these two terms with their respective probabilities:
\begin{equation} 
\begin{split}
H' &= p(1)H(1) + p(-1)H(-1)\\
       &\approx H_{0} -\frac{V[\langle \mathcal{O} \rangle]}{2\text{ln}2}-\frac{E[\langle \mathcal{O} \rangle]^2}{2\text{ln}2}(1-\frac{2}{N}).
\end{split}
\end{equation}

Therefore, the expected information gain after measuring $\mathcal{O}$ is:

\begin{equation} \label{eq:Iexp}
\begin{split}
I &= H_{0} - H'\\
&\approx H_{0} - H_{0} + \frac{V}{2\text{ln}2}+ \frac{E^{2}}{2\text{ln}2}(1-\frac{2}{N})\\
&= \frac{V}{2\text{ln}2}+ \frac{E^{2}}{2\text{ln}2}(1-\frac{2}{N}).
\end{split}
\end{equation}

If observable $\mathcal{O}$ is traceless, we use Equation~\ref{eq:E} to conclude that $E = 0$ and finally obtain:

\begin{equation} \label{eq:Iexp2}
\begin{split}
I &\approx \frac{V}{2\text{ln}2}.
\end{split}
\end{equation}

\nocite{*}

\twocolumngrid
\bibliography{apssamp}

\end{document}